# Nuclear Magnetic Resonance Diffusometry of Linear and Branched Wormlike Micelles

Samuel W. Holder,[1,2] Samuel C. Grant[1,2] and Hadi Mohammadigoushki[1,2] *

[1]*Department of Chemical & Biomedical Engineering, FAMU-FSU College of Engineering, Tallahassee, FL, 32310, United States*

[2]*Center for Interdisciplinary Magnetic Resonance, National High Magnetic Field Laboratory, Florida State University, Tallahassee, FL, USA 32310*

Diffusion studies using nuclear magnetic resonance (NMR) spectroscopy were conducted on two model surfactant solutions of cetyltrimethylammonium bromide/sodium salicylate (CTAB/NaSal) and cetylpyridinium chloride/sodium salicylate (CPCl/NaSal). By increasing the salt-to-surfactant concentration ratio, these systems display two peaks in the zero-shear viscosity and relaxation time, which are indicative of transitions from linear to branched micellar networks. The goal of this work is to assess the sensitivity of NMR diffusometry to different types of micellar microstructures and identify the mechanism(s) of surfactant self-diffusion in micellar solutions. At low salt-to-surfactant concentration ratios, for which wormlike micelles are linear, the surfactant self-diffusion is best described by a mean squared displacement, $Z^2$, that varies as $Z^2 \propto T_{diff}^{0.5}$, where $T_{diff}$ is the diffusion time. As the salt concentration increases to establish branched micelles, $Z^2 \propto T_{diff}$, indicating a Brownian-like self-diffusion of surfactant molecules in branched micelles. This result indicates that NMR diffusometry is capable of differentiating various types of micellar microstructure. In addition, the self-diffusion coefficient of the surfactant molecules in linear and branched micelles are determined, for the first time, by comparing to the existing restricted diffusion models, and shown to be much slower than the diffusion of proton molecules in the bulk. Moreover, in linear and moderately branched wormlike micelles, the dominant mechanism of surfactant self-diffusion is through the curvilinear diffusion of the surfactant molecules along the contour length of the micelles, whereas in the branched micelles, before the second viscosity maxima, the surfactant self-diffusion could arise from a combination of micellar breakage, exchange between micelles and/or the bulk.

*Corresponding author: hadi.moham@eng.famu.fsu.edu

# I. INTRODUCTION

Surfactants or surface-active agents are amphiphilic molecules that have received immense scientific and technological interest. When mixed in aqueous solutions, these molecules self-assemble into a variety of interesting and highly dynamic nanostructures including spherical, cylindrical, vesicle or wormlike [1, 2]. The shape of the micelles depend on the surfactant packing parameter, p, which is the ratio of the surfactant hydrophobic tail area to the hydrophilic head area [3, 4]. At a constant surfactant concentration, the addition of salt transforms spherical micelles to isolated rod-like micelles, which at sufficiently high concentrations of salt and surfactant leads to growth in micellar length well beyond the persistence length towards entangled wormlike structures for p $\leq$ 1/2.

The addition of salt to the surfactant solution causes an increase in the rheological properties (*e.g.*, the zero-shear rate viscosity $\eta_0$ or the shear relaxation time $\tau_R$), which is mainly due to the change in micellar topology from spherical to rod-like to entangled linear wormlike structures. While entangled wormlike micellar solutions (WLMs) share many similarities with entangled polymer solutions, there is a distinct difference between these two class of materials in that wormlike micelles can break and reform. Therefore, in addition to reptation, micelles can relax via breakage and reformation mechanisms. According to Cates and co-worker [5], in the fast-breaking regime for which micellar breakage time ($\tau_{br}$) is much shorter than the reptation time ($\tau_{rep}$), the linear viscoelastic rheology of the WLMs is best described by a single-mode Maxwell model within a wide range of frequencies. Yesilata and co-workers [6] showed that, in the fast-breaking regime, the breakage time of the micelles can be estimated as the inverse of the frequency at which the loss modulus shows a local minimum.

Further increase of the salt concentration at a constant surfactant concentration results in appearance of a maximum in the zero-shear rate viscosity and/or the micellar relaxation time in a wide range of systems including cetyltrimethylammonium bromide/sodium salicylate (CTAB/NaSal), cetylpyridinium chloride/sodium salicylate (CPCl/NaSal), and erucyl bis(hydroxyethyl)methylammonium chloride/potassium chloride (EHAC/KCl) as well as other systems [7, 8, 9, 10, 11, 12, 13, 14, 15]. At even higher salt concentrations, a second maximum is documented for micellar systems of CTAB/NaSal and CPCl/NaSal [13, 16, 17]. The microstructural origin of the zero-shear-rate viscosity or relaxation peaks has been the main focus of several research studies over the past couple of decades [7, 8, 9, 10, 11, 12, 13, 14, 15, 18]. It has emerged unambiguously from direct cryogenic Transmission Electron Microscopy (cryo-TEM) imaging that the first maximum in a wide range of WLMs is linked to a microstructural transition from linear wormlike micelles to branched micellar structures [9, 19, 20, 15]. Additionally, theoretical studies of Lequeux [21] have predicted that micellar branching should lower the viscosity, fundamentally supporting the transition from linear to branched micelles. Beyond the first viscosity peak, screening of the electrostatic interaction favors formation of the branched micelles. Despite ample evidence of the microstructural origin of the first viscosity maximum in many micellar systems, the direct cryo-TEM imaging of the micellar solutions around the second viscosity peak is not available. Therefore, it is not exactly clear what gives rise to a second viscosity peak in some micellar solutions. Using the diffusive wave spectroscopy technique, Oelschlaeger et al. [13] suggested that the viscosity increase before the second viscosity peak could be due to an increase in micelles length and a reduction in branching density, while beyond the second viscosity maxima, the viscosity drop is due to a decrease in micelles length and an increase in the branch density [13]. However, this suggestion relies on an approximate formula that does not take into account the effects of micellar branching (see more details below on section II.A). Although cryo-TEM imaging is crucial for confirming the nature of microstructural transitions in WLMs, this method is extremely challenging from sample preparation to the image acquisition stage, with the equilibrium microstructure of the micelles often altered during sample preparation [22]. Therefore, recent work has sought alternative techniques that could be sensitive to the type of micellar microstructure (*i.e.*, linear vs. branched). As an example, mechanical tools such as extensional rheometry [23, 24, 17], scattering techniques such as Neutron Spin Echo (NSE) and Dynamic Light Scattering (DLS) [25] and Nuclear Magnetic Resonance (NMR) [26] have proven to be helpful for this purpose.

Recently, Calabrese and Wagner used NSE in combination with DLS techniques to study the transition from linear to branched structures in WLMs of cetyltrimethylammonium tosylate/sodium dodecyl benzene sulfonate (CTAT/SDBS/NaToS) in $D_2O$ [25]. These researchers showed that on the scale of micelle's segment, the measured normalized intermediate neutron scattering function for linear micelles is consistently different from branched micelles, which could be used as a criterion to distinguish these micelles [25]. On the other hand, extensional experiments of Omidvar et al. [24], Sachsenheimer et al. [17] and Chellamuthu & Rothstein [23] suggest that the response of branched and linear WLMs is different in uniaxial extensional flows, and therefore, extensional flows are sensitive to the type of micellar microstructures. A more detailed discussion on the effects of micelles microstructure on flows of wormlike micelles is provided in a recent review [27]. In addition, Angelico et al. [26] showed that Pulsed Gradient Spin Echo (PGSE) NMR can distinguish branched micelles from linear micelles in a reverse micellar solution based on lecithin in a mixture of isooctane ($iC_8$) and cyclohexane ($cC_6$). Angelico et al. showed that, in the linear reverse micelles, the mean squared displacement (MSD) varied as the root square of the diffusion time, while in the branched networks, MSD was linearly dependent on the diffusion time [26]. We note that the latter study is limited to a reverse micellar system, while the majority of prior microstructural investigations have been carried out on micellar systems for which continuous phase is water or deuterium oxide ($D_2O$). In reverse micelles based on lecithin, the continuous phase is oil, and therefore, electrostatic interactions are negligible, while in aqueous-based WLMs, electrostatic interactions are present. Moreover, in the reverse branched micelles used by Angelico et al. [26], the micelles breakage is not significant, because their linear viscoelastic results do not fit to a single-mode Maxwell model. However, previously published work (and this study) on branched micelles show that the linear rheology of some well-studied branched micellar systems is best described by a single-mode Maxwell model over a wide range of frequencies. Hence, in contrast to the reverse micelles of lecithin in isooctane/cyclohexane, the micellar breakage could be an important factor in the branched micellar solutions that are made in the aqueous phase, which consequently may affect the NMR measurements and their sensitivity to the linear-branched micellar transition in WLMs. In addition, as pointed out before, many wormlike micellar solutions exhibit a second peak in the zero-shear rate viscosity at high salt-to-surfactant concentration ratios, a feature that is not present in the reverse micellar solution studied previously. Therefore, it is still an open question as to whether NMR is capable of distinguishing different micellar microstructures (*i.e.*, linear and branched). Moreover, the mechanism(s) of surfactant self-diffusion in the linear and branched micelles are not well understood (see Section II below). Hence, NMR-based diffusion measurements on WLMs deserve further investigation.

The main goal of this work is twofold: First, we investigate whether diffusion-weighted NMR spectroscopy is sensitive to the type of wormlike micellar structure (*i.e.*, linear vs. branched). Second, we directly identify the mechanism(s) of surfactant self-diffusion in entangled WLMs by evaluating the surfactant self-diffusion coefficients for a range of microstructural regimes. These experiments are carried out over a broad range of salt-to-surfactant concentration ratios covering both viscosity maxima. For this purpose, two well-studied micellar solutions based on CPCl/NaSal and CTAB/NaSal that show two viscosity peaks reminiscent of transitions from linear to branched structures were selected. For a similar micellar solution based on CPCl/NaSal, the direct evidence for transition from linear to branched micelles is documented by Gaudino et al. [14]. Although similar systematic TEM imaging is not available for the CTAB/NaSal solution, Francisco et al. [28] showed through TEM that micelles in the solution of CTAB/NaSal (100 mM/100 mM) form branched structures. This composition corresponds to the salt concentrations beyond the first viscosity maximum in our experiments (see details below). Additionally, extensional rheological results of Sachsenheimer et al. [17] showed that CTAB/NaSal system behaves similarly to the CPCl/NaSal solution for salt concentrations beyond the first viscosity maximum. Therefore, based on the above, the micellar solution of CTAB/NaSal should experience a similar linear to branched micellar transition beyond the first viscosity maximum. This hypothesis will be tested by cross comparing the self-diffusion results of CTAB/NaSal and CPCl/NaSal solutions.

## II. POTENTIAL MECHANISMS OF MOLECULAR DIFFUSION IN WLMs

### A. Micellar Curvilinear Diffusion due to Reptation and/or Breakage-Reformation:

Cylindrical micelles much longer than their persistence length $l_p$ are flexible and behave similarly to flexible polymers. However, as noted before, unlike polymers, wormlike chains can break and reform. As a result, micellar curvilinear motion is affected by a combination of reptation and breakage/reformation. To evaluate the relative importance of each of these mechanisms, Cates [29] has introduced a dimensionless time scale, $\xi = \tau_{br}/\tau_{rep}$, for which $\tau_{br}$ and $\tau_{rep}$ denote the breakup time (or "lifetime" of the micelles) and micellar reptation time, respectively. For $\xi > 1$, micelles diffuse similar to polymers mainly through a reptation process, in which the apparent diffusion coefficient of the micelles is described by the following relation [30]:

$$D_{mic} = D_{rep} = \frac{k_B T}{6\pi\eta\,\zeta}(\frac{L_e}{L_c})^2,$$  (1)

where $k_B$, T, $\eta$, $\zeta$, $L_e$ and $L_c$ are the Boltzmann constant, temperature, viscosity, network mesh size, entangled length and contour-length of the micelles, respectively.

On the other hand, in the fast-breaking regime (i.e., $\xi \ll 1$), a portion of the micelle diffuses over a length $X$ before breaking, and then the tube renews due to micelle reformation. Thus, the length $X$ is the distance that the micelle diffuses curvilinearly before chain scission occurs on this part of the tube. According to Schmitt and Lequeux [30], in the fast-breaking regime, the curvilinear diffusion of the micelles can be approximated as:

$$D_{mic} \approx D_{rep}\xi^{-1/3}.$$  (2)

The above relation indicates that the apparent diffusion coefficient (ADC) of the micelles will be enhanced as the micellar breakage rate is strengthened (or equivalently as $\xi$ decreases). Therefore, the relative importance of the micellar breakage and reptation on micellar curvilinear diffusion mechanisms can be evaluated through the dimensionless breakage time ratio $\xi$.

## B. Surfactant self-diffusion in Micellar Solutions:

### 1. Surfactant Exchange Between Micelles and/or the Continuous Phase:

Separate from micellar curvilinear diffusion due to breakage/reformation or reptation, if the residence time of the individual surfactant molecules on the micelles is shorter than $\tau_{br}$ or $\tau_{rep}$, surfactant molecules may be exchanged between micelles or between a micelle and the aqueous bulk. Exchange of the surfactant between micelles is possible solely at the entanglement points. If this exchange is the primary mechanism of the surfactant diffusion, the diffusion due to entanglement, $D_{ent}$, is directly related to micellar entanglement density as $D_{ent} \propto N$, where $N$ denotes the average number of entanglement points per unit length of the micelle. The significance of the surfactant exchange between micelles at entanglement points can be evaluated in experiments by monitoring the variation of the ADC as a function of the micellar entanglement density.

On the other hand, surfactants can be exchanged between micelles and the bulk. As the micellar network grows denser and more interconnected, the impact of the micelle-bulk surfactant exchange weakens due to the diminished contact with the bulk. Therefore, the diffusion coefficient due to exchange between the bulk and the micelle, $D_{exc}$ is inversely proportional to the micelle concentration as $D_{exc} \propto 1/C_m$, where $C_m$ denotes the concentration of the micelles in the medium.

Moreover, this diffusion mechanism will give rise to a MSD that is linearly proportional to the diffusion time (*i.e.*, random-like diffusion). The importance of this diffusion mechanism is assessed in experiments by monitoring the diffusion coefficient as a function of micelle concentration and/or the variation of MSD with respect to the diffusion time.

### 2. Surfactant Curvilinear Diffusion in Micelles:

In addition to the above mechanisms, surfactant molecules, due to their small size, are able to diffuse rapidly along and inside a wormlike micelle's longest dimension (contour length). Referred to as surfactant curvilinear diffusion, this mechanism should be much faster than the diffusion of the micelles themselves because surfactant molecules are much smaller than the wormlike micelles. The significance of this diffusion mechanism can be assessed by evaluating how the MSD varies with molecular diffusion time, and also by the quality of the fit to the theoretically developed models for the curvilinear diffusion (see below).

## III. MOLECULAR DIFFUSION IN THEORY

Although multiple mechanisms could contribute into surfactant diffusion in the micellar solutions, the relevant theoretical models are limited to the case of free self-diffusion (due to Brownian motion) and the curvilinear diffusion inside the micellar tubes.

In a seminal work, Stesjkal and Tanner [31] showed that Brownian diffusion of small molecules in an isotropic medium could be linked to attenuation of their NMR signal as the following:

$$\frac{S(q)}{S(0)} = \exp[-D\,q^2 T_{diff}], \tag{3}$$

where $S(q)$ is the NMR signal as a function of diffusion weighting q, $S(0)$ is the NMR signal in the absence of diffusion weighting (q = 0), and D is the apparent diffusion coefficient. q is defined as $q = \gamma\delta g$, where $\gamma$ is the gyromagnetic ratio of the probe (*e.g.*, for $\gamma_{1_H} = 2.67 \times 10^8$ rad/s/T), and g is the magnitude of the magnetic gradient pulse. In addition, $T_{diff} = (\Delta - \delta/3)$, where $\Delta$ is the diffusion time and $\delta$ is the diffusion gradient time. For this type of free self-diffusion, the MSD, $Z^2(T_{diff})$, varies linearly with the diffusion time and is equal to: $Z^2(T_{diff}) = 2DT_{diff}$.

In the case of surfactant curvilinear diffusion inside a reverse wormlike micelle, Angelico et al. [32] developed the following relation:

$$\frac{S(q)}{S(0)} = \exp(\Gamma^2)\,erfc(\Gamma)\,, \tag{4}$$

where $\Gamma = \frac{(D_c \, T_{diff})^{1/2} \, \lambda q^2}{3}$. $D_c$ is the apparent curvilinear self-diffusion coefficient of the surfactant along the micellar tube length, and $\lambda$ is a characteristic diffusion step length. Angelico et al. [32] hypothesized that $\lambda$ is related to the micelle persistence length ($l_p$) or micellar radius of gyration.

Tanner and Stesjkal [31] developed a model for the restricted diffusion between two infinite parallel plates as:

$$\frac{S(q)}{S(0)} = \left[ \frac{2[1 - \cos(2\pi q L_z)]}{(2\pi q L_z)^2} + 4(2\pi q L_z)^2 \times \sum_{n=1}^{N} \exp\left(-\frac{n^2 \pi^2 D_c \Delta}{L_z^2}\right) \times \frac{1 - (-1)^n \cos(2\pi q L_z)}{\left((2\pi q L_z)^2 - (n\pi)^2\right)^2} \right] \quad (5)$$

In the above equation, $L_z$ denotes the distance between the plates. Although simplistic, this configuration could be a possible first principle approximation of the micellar shape as has been applied to evaluate restricted macromolecular diffusion in mixed micellar solutions [33]. Therefore, this model will be used as a simple confining geometry to evaluate the restricted self-diffusion of surfactants in wormlike micelles.

Perhaps a more accurate potential estimation of the micellar geometry is a cylindrical tube. Callaghan and co-workers [35] developed a model for 1D curvilinear molecular self-diffusion of small molecules along the longest axis of randomly oriented cylindrical capillaries as:

$$\frac{S(q)}{S(0)} = \int_0^1 \exp(-q^2 D \, T_{diff} \, x^2) \, dx \, . \quad (6)$$

Furthermore, these authors expanded the above equation to account for a 2D restricted self-diffusion which reads as [35]:

$$\frac{S(q)}{S(0)} = \exp(-q^2 D \, T_{diff}) \int_0^1 \exp(-q^2 D \, T_{diff} \, x^2) \, dx \, . \quad (7)$$

Although they can diffuse curvilinearly along the micellar length, surfactants can self-diffuse through branches as those branches start to form. In the framework of the wormlike micelle chain, this restricted self-diffusion may be approximated by a 2D restricted self-diffusion. For a three-dimensional unrestricted self-diffusion of molecule, the above equations (Eqs. 6-7) simplifies to the familiar Eq. (3). The curvilinear diffusion of the surfactant in the micellar tube is restricted and the above relations (Eqs. 4-7) can be fitted to the experimental data to obtain the curvilinear ADC of the individual surfactants along the contour length of the wormlike micelles.

## IV. MATERIALS

Two model wormlike micellar solutions were evaluated in this study. The first system contains CPCl and NaSal in 99% deuterated water ($D_2O$) and the second system consists of CTAB and NaSal in 99% $D_2O$. CTAB was purchased from Spectrum Chemicals, CPCl and NaSal were purchased from Sigma-Aldrich and used as received. Deuterated water was purchased from Cambridge Isotope Laboratories and used as received. Bulk solutions were made at fixed surfactant concentrations of 100 mM for both CPCl and CTAB. The ratio of the salt-to-surfactant concentration R was varied as R=[NaSal]/[CPCl]=0.45–4 for CPCl/NaSal and R = [NaSal]/[CTAB] = 0.3–4 for CTAB/NaSal solutions.

## V. EXPERIMENTAL

### A. Fluid Characterization

To characterize the rheology of the wormlike micellar fluids, a commercial rheometer (Anton-Paar model MCR-302) was used. A standard cone-and-plate geometry with a cone diameter of 50 mm and cone angle of 1° was used for these measurements. The small amplitude oscillatory shear (SAOS) experiments were performed to obtain the longest relaxation time of the micellar solutions, micellar entanglement density and the average contour-length of the wormlike micellar solutions. Frequency sweep was completed at T = 20.0±0.5 °C. In addition, steady shear experiments were carried out at the same temperature over a wide range of imposed shear rates ($10^{-3}$-$10^{2}$ s$^{-1}$), to determine the zero-shear-rate viscosity of the micellar solutions.

### B. Molecular Diffusion Measurements

To evaluate the diffusion behavior of the surfactants in WLMs, samples were analyzed using an 11.75-T, widebore magnet equipped with a 500-MHz NMR spectrometer (AVI, Bruker Biospin, Billerica, MA) and microimaging apparatus (Micro2.5 Bruker gradient system) capable of achieving 1 T/m gradients on three axis. Samples were analyzed in sealed 5-mm NMR tubes. Data were acquired using a 5-mm diameter, $^1$H linear birdcage at T = 20.0±0.5 °C. Diffusion-weighted (DW) $^1$H spectra at 500 MHz were acquired using a stimulated echo acquisition mode (DW-STEAM) sequence (see Fig. 1). For these experiments, the echo time (TE) and repetition time (TR) were kept constant at 16 and 2000 ms, respectively. Protons ($^1$H) was used as the NMR probe to increase the overall NMR signal-to-noise ratio. Diffusion weighting was imparted by the application of a pulse gradient pair, located before the 2$^{nd}$ 90° RF pulse and the last 90° RF pulse, so that the mixing time (TM) between these pulses could be used to alter the gradient separation time ($\Delta$) without impacting T$_2$ relaxation of the acquired signal. To characterize the diffusion profile, eight $\Delta$ times (16, 21, 51, 101, 201, 301, 401 and 501 ms) were acquired in separate experiments each with 16 different values of gradient magnetic field strength ($g$ = 0-754.6 mT/m) with a fixed gradient pulse duration ($\delta$) of 3 ms. These experiments covered a range of diffusion times (T$_{diff}$) from 15-500 ms as well as q values from 0-0.095 μm$^{-1}$. The parameter q is linked to the probability distribution of the diffusion displacement [36]. In typical NMR experiments, this length scale (or inverse of q) is selected to be in the order of ~ 10 μm (or larger) to monitor the translational diffusion of various molecules with different dynamics (or sizes). Fundamentally, the NMR technique identifies the chemical shifts within a sample based on the atomic structure and interactions regardless of the molecular size. Once chemical resonances are identified, NMR can probe molecular motion with sensitivity to either molecular restrictions or anisotropy within the media. In this study and in an otherwise isotropic medium, the well-established DW-STEAM sequence is used to probe the molecular restrictions experienced by surfactants in WLMs through NMR signal decay over a range of different diffusion weightings. To vary the diffusion weighting q, the gradient field magnetic strength, g, was incremented.

## C.    Postprocessing and Fitting

[1]H DW-STEAM datasets were processed in the Bruker TopSpin 4.0.4 software. Time-domain data were apodized using a 10-Hz exponential filter prior to 1D Fourier transform. Spectra were phase-corrected, and analysis was performed on spectral components using both peak intensity and peak integrals as a function of diffusion weighting for each diffusion time and direction. The NMR signal data were fitted to the model equations (Eqs. 3-7) described above in MATLAB 2017b (MathWorks, Natick, MA) using a conventional nonlinear regression technique (*fitnlm*).

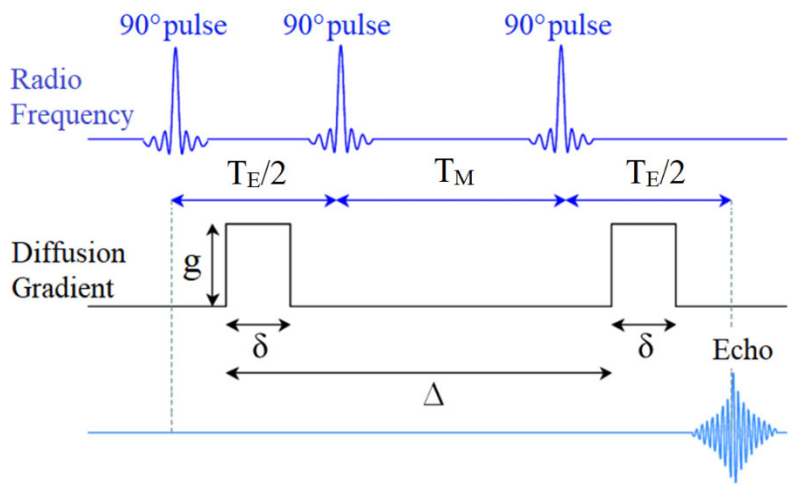

**FIGURE 1.** A schematic showing a diffusion weighted stimulated pulse gradient spin echo (DW-STEAM) protocol used in this paper.

## VI.    RESULTS AND DISCUSSION
## A.    Rheology

Fig. 2a shows the zero-shear-rate viscosity of the micellar solutions as a function of salt to surfactant concentration ratio R for the two wormlike micellar solutions. For both systems, the zero-shear-rate viscosity $\eta_0$ increases as R increases up to a critical value at which $\eta_0$ shows a maximum. For the system of CPCl/NaSal, the maximum of the zero-shear rate viscosity occurs at R=0.6, while for the CTAB/NaSal system, it is in the vicinity of R=0.4. Further increase in the salt concentration beyond the first viscosity peak leads to a drop in the zero-shear rate viscosity, which is followed by a second viscosity maximum at R=2 for both CPCl/NaSal and CTAB/NaSal systems. These results are consistent with the existing literature [13, 37].

It is worth noting that the zero-shear-rate viscosity at the second peak in the CPCl/NaSal system is comparable to the first viscosity peak (see Table S(1) in the supplementary materials for numerical values). Typically, in aqueous solutions of CPCl/NaSal with deionized water, the second viscosity peak is much lower than the first viscosity peak [13]. This difference is presumably linked to the type of solvent used in this work. It has been shown recently that $D_2O$ alters the rheology and microstructure of the micellar systems when used as an aqueous medium [38].

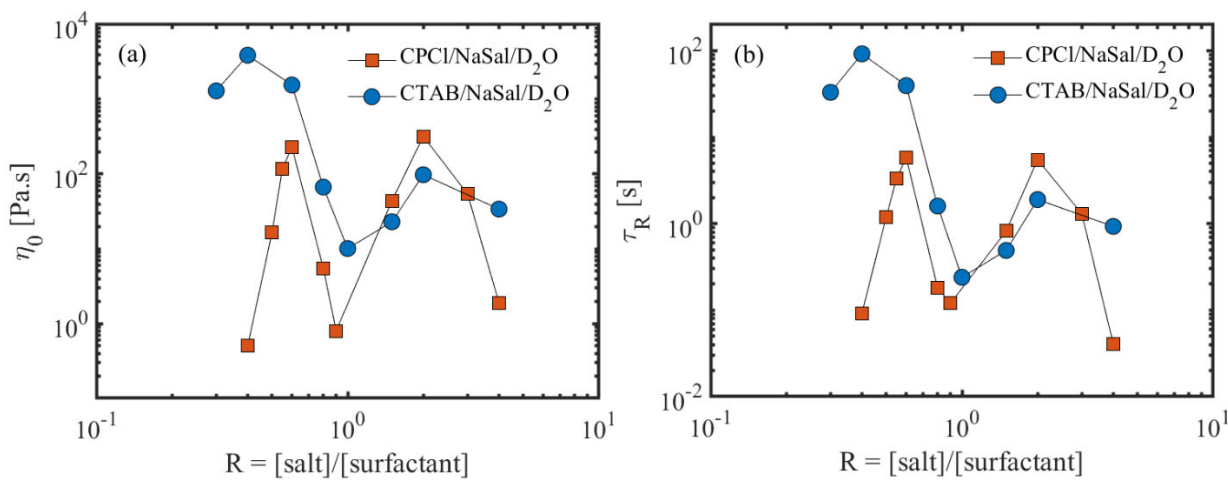

**FIGURE 2.** (a) Zero shear-rate viscosity and (b) the longest relaxation time of the micellar solutions as a function of salt to surfactant concentration ratio for the deuterated micellar solutions used in this study.

Additionally, Fig. 2b shows the variation of the longest micellar relaxation time $\tau_R$ as a function of salt-to-surfactant concentration ratio for these solutions. The relaxation time is obtained by fitting a Maxwell model to dynamic moduli ($G^{'}$ and $G^{''}$ as a function of angular frequency) data (see Fig. S1 in the supplementary materials). Notably, the majority of these systems follow a single-mode Maxwell model within a wide range of frequencies. For those solutions that are best described by a single-mode Maxwell model, we have calculated the breakage time ratio (see Table (1) and supplementary materials), which will be used to evaluate the relative importance of micellar breakage to that of the reptation.

In addition to the relaxation time, the results of small amplitude oscillatory shear (SAOS) experiments allow us to estimate some of the microstructural properties of the wormlike micelles including micellar contour-length and the micellar entanglement density.

As introduced by Granek and Cates [39], and later modified by Granek [40], the entanglement density of the micelles in the fast-breaking regime (i.e., $\xi << 1$) can be linked to the linear viscoelastic properties as:

$N = \frac{L_c}{L_e} \cong (\frac{G_o}{G''_{min}})^{5/4},$ (8)

where, $N$, $L_c$, $L_e$, $G_0$ and $G''_{min}$ denote the entanglement density of the micelles, micellar contour-length, average micelles entanglement length, plateau modulus and the local minimum in the loss modulus at high frequencies, respectively. In addition, the plateau modulus is linked to microstructural properties via: $G_o \cong \frac{k_B T}{L_e^{9/5} l_p^{6/5}},$ (9) where $k_B$ denotes Boltzmann's constant.

Using Eqs. (6-7) in conjunction with SAOS measurements, the average entanglement density and the contour length of the micelles in these WLMs can be calculated. For those solutions in which $G''_{min}$ occurs at frequencies beyond the accessible range of the rheometer ($\omega > 100$ rad/s), $N$ and $L_c$ values are extracted from the published literature data. In such instances, we have used reported values for $G_o$ and $G''_{min}$ in combination with the Eqs. (8-9) to obtain the micelles entanglement density and the average micellar length. In calculating the micellar length in these systems, persistence length of the CPCl/NaSal and CTAB/NaSal solutions are used from diffusive wave spectroscopy measurements reported in the literature [13, 37]. Table (1) shows the list of micellar contour length and the entanglement density for the surfactant solutions used in this work. These properties are crucial in assessing the mechanism(s) of the surfactant self-diffusion in the micellar systems.

It is worth noting that the approximate relation proposed by Granek does not take into account the effects of micellar branching. In branched micelles, inter-micellar branching distance may become important in addition to entanglements. Additionally, the method of defining a micelle counter-length in a branched micelle is still unknown. Even in the presence of charged interactions, these equations provide an approximate value for the micellar entanglement density. Therefore, to determine more realistically the microstructural properties of the micelles in the branched regime, further improvement of the above theoretical studies are necessary.

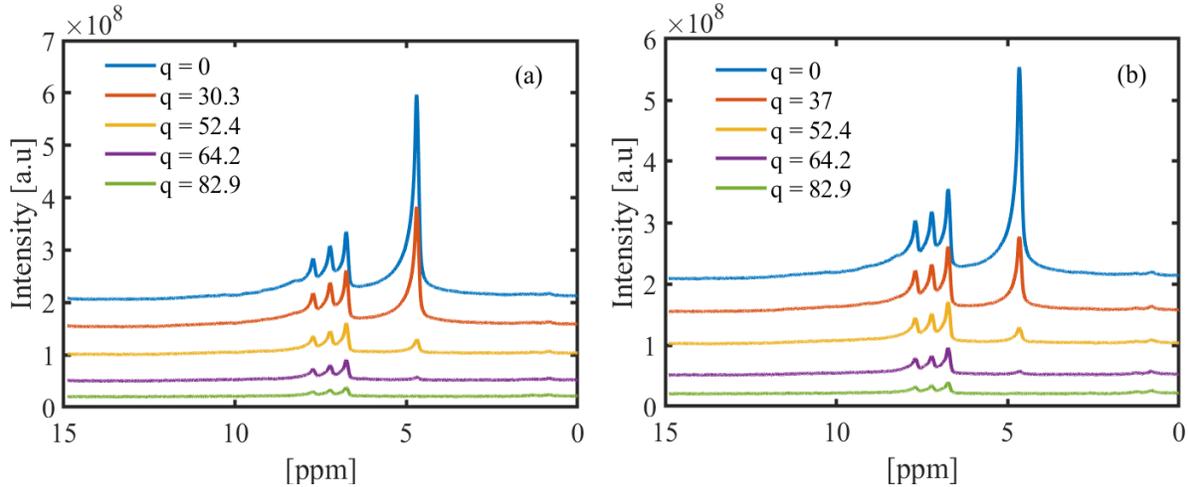

**FIGURE 3.** Raw NMR signal intensities as a function of chemical shift for (a) CPCl/NaSal at R = 3 and (b) CTAB/NaSal at R = 1.5. These data are acquired by imposing a diffusion time $T_{diff}$ = 200 ms. The unit for q is [mm⁻¹].

## B. Surfactant Self-diffusion via Diffusion Weighted NMR Spectroscopy

In this section, DW-STEAM data and model fitting are assessed mainly 1) to probe whether diffusion weighted NMR spectroscopy can be used to distinguish linear from branched micellar networks and 2) to identify the dominant mechanism(s) of surfactant self-diffusion in micellar solutions with different microstructures. Fig. 3 shows a collection of phase-corrected diffusion weighted $^1$H spectra over a broad range of chemical shifts [in ppm] for (a) CPCl/NaSal/$D_2O$, R = 3 and (b) CTAB/NaSal/$D_2O$, R = 1.5 at different q values. In both systems, the largest peak at a chemical shift of 4.7 ppm corresponds to the residual protons available in the otherwise deuterated media. In the CPCl/NaSal system, four additional peaks are observed at chemical shifts of 0.75, 6.7, 7.3 and 7.7 ppm. The peak at 0.75 ppm belongs to ω-$CH_3$ in the CPCl surfactant and three peaks at 6.7, 7.3 and 7.7 ppm correspond to the salicylate ion [41, 42]. On the other hand, in the CTAB/NaSal/$D_2O$ system (Fig. 3b), five peaks are observed. The peaks at 0.75 ppm and 1.3 ppm correspond to the ω-$CH_3$ and $(CH_2)_n$ in the backbone of the CTAB, respectively [43]. The remaining three peaks at 6.7, 7.3 and 7.7 ppm correlate with the salicylate ion [43, 42]. In both micellar systems, across all R values, the peak associated with the residual proton in the $D_2O$ medium (*i.e.*, 4.7 ppm) decays sharply (and mono-exponentially) as a function of diffusion weighting q, dropping to noise at high diffusion weighting (see Fig. 4(a)). This mono-exponential decay is fitted to Eq. (3), and the ADC of the residual protons in the $D_2O$ medium was extracted for all WLMs. Fig. 4(b) shows the ADC of the proton in the $D_2O$ across all WLMs. In addition, we measured the ADC of the residual protons in pure $D_2O$ to be about $1.79 \times 10^{-9}$ (m$^2$/s) (The dashed line in Fig. 4(b)). Fig. 4(b) indicates that the proton self-diffusion in the micellar solutions at different salt-to-surfactant concentrations is slower than its self-diffusion in otherwise pure $D_2O$. The reduction in proton self-diffusion in WLMs is presumably caused by two factors. First, presence of the surfactant micelles could obstruct the diffusion of protons in the bulk $D_2O$. Second, a fraction of protons in the $D_2O$ formulations is bound to micelles due to hydration. These two factors cause the proton to have a lower mobility in the micellar solutions. Similar result has been reported in the self-diffusion of proton in non-ionic surfactant solutions [44].

In contrast to residual protons in the $D_2O$ medium, the signals associated with surfactant molecules generally show a much more gradual decay. This trend implies a restriction of the surfactant self-diffusion as opposed to the rapid signal decay seen in the solvent. To determine the mechanism of the restricted surfactant diffusion in the micellar systems, the NMR signal decay for surfactant peaks were calculated in relation to $T_{diff}$ and q. The first part of this section is devoted to evaluation of the sensitivity of the NMR diffusion measurements to the type of wormlike micellar microstructure. To this end, the attenuated NMR signal can be expressed as:

$$S(q)=S(0) \ \exp\left[-\frac{1}{2}Z^2(T_{diff})q^2\right], \tag{10}$$

where $Z^2(T_{diff})$ denotes the mean squared displacement of the tracked surfactants, which can be approximated regardless of the diffusive behavior as:

$$Z^2(T_{diff}) \propto T_{diff}^{\alpha}, \tag{11}$$

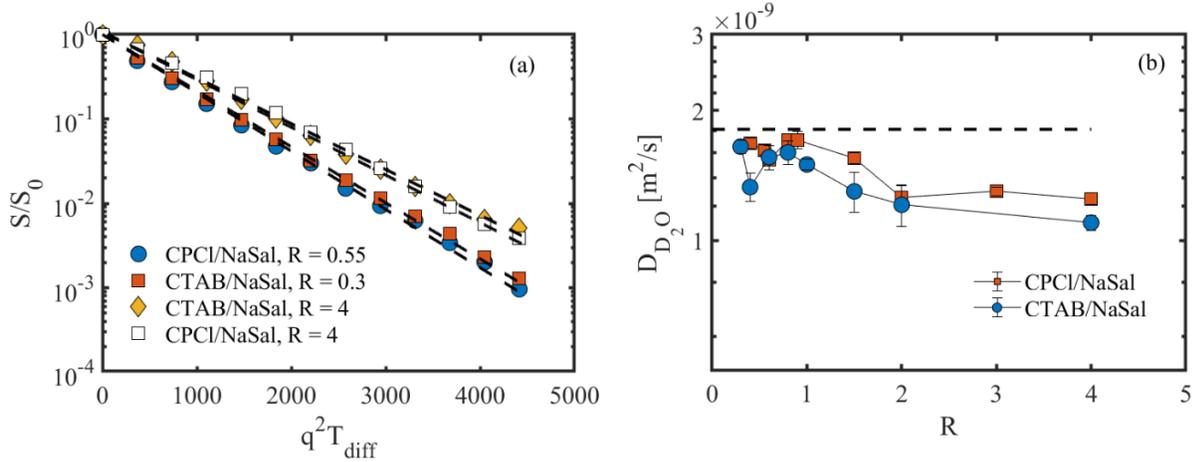

**FIGURE 4.** (a) Normalized $D_2O$ NMR signal intensity as a function of diffusion weighting for sample wormlike micellar solutions at $T_{diff} = 50$ ms. (b) Diffusion coefficient of the proton in the bulk $D_2O$ of the surfactant solutions. Dashed lines in (a) represent mono-exponential decay (Eq. (1)). The dashed-line in part (b) corresponds to the diffusion of proton in the pure $D_2O$ ($1.79 \times 10^{-9}$ ($m^2$/s)).

for which $\alpha$ is a power-law index that can be linked to the fractal dimensionality of the micellar solution and key values of $\alpha$ can be used to describe specific diffusive behaviors. $\alpha$ is inversely proportional to the diffusion restriction in a sample. Free and unrestricted self-diffusion of molecules in the bulk is characterized by a linear relationship between $Z^2(T_{diff})$ and $T_{diff}$, with $\alpha=1$. As restriction increases, $\alpha$ decreases approaching a key value at $\alpha=1/2$, which characterizes curvilinear diffusion of molecules in tubes such as that seen in Eqs. (4-5). It is possible for $\alpha$ to increase as well, to the established key value of $\alpha=3/2$, characterizing a form of "super-diffusion" modeled via a combination of diffusive and ballistic formulae [45]. However, while the latter diffusive behavior has been previously found in reverse WLMs [45], it is not observed in experiments with CPCl/NaSal and CTAB/NaSal systems used in this study (see details below). In order to determine the correct value of $\alpha$ for these WLMs at different salt to surfactant concentration ratios, the dimensionless NMR signal measured in experiments was be plotted against $q^2T^{\alpha}_{diff}$. With the correct value of $\alpha$, all experimental data will collapse into a single curve.

Fig. 5 shows the normalized NMR intensity signal as a function of $q^2 T_{diff}^{\alpha}$ for the wormlike micellar systems of CPCl/NaSal/$D_2O$ (top row) and CTAB/NaSal/$D_2O$ (bottom row) at various salt-to-surfactant concentration ratios, R. The relationship between the mean squared displacement $Z^2(T_{diff})$ and $T_{diff}$ is displayed on the x-axis of each subfigure. At low R, before the first viscosity maxima (*i.e.*, Fig. 5(a,d)), the experimentally measured dimensionless NMR signals at different diffusion times are superimposed each other with $\alpha \approx 0.55 \pm 0.03$. As R increases beyond the first viscosity maxima (Fig. 5(b,e)), $\alpha$ increases to $\alpha \approx 0.8 \pm 0.1$ for R = 0.9 and eventually approaches linearity until the second viscosity maxima for both systems are reached at R = 2 (see Fig. 5(c,f)).

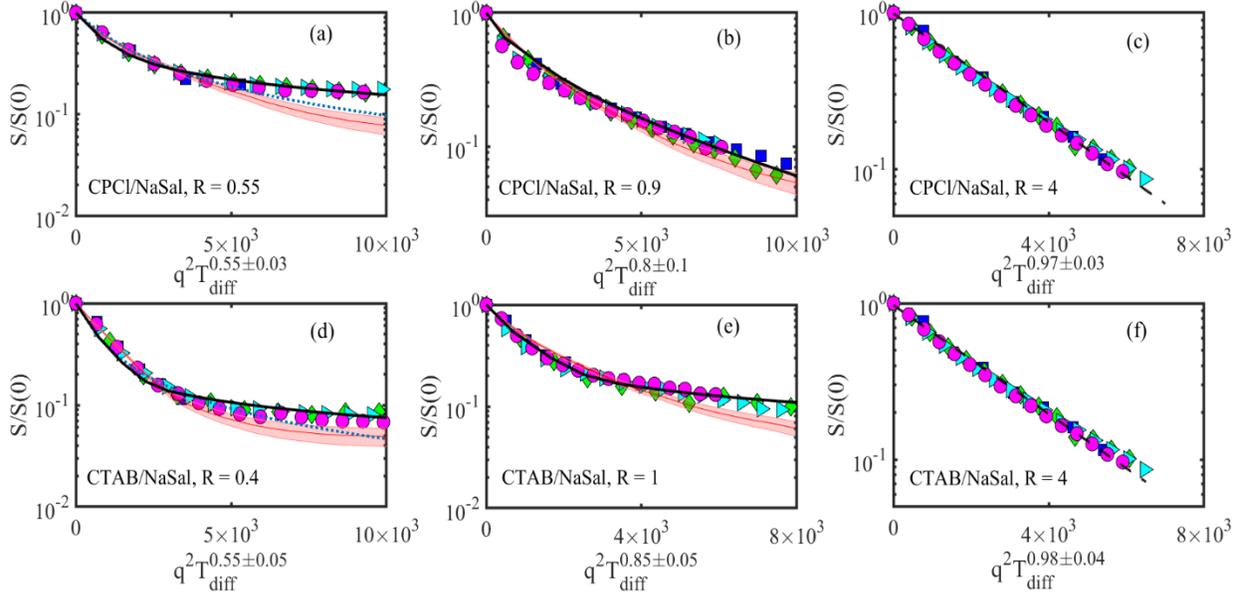

**FIGURE 5.** Normalized NMR signal intensity of the surfactant peak (at a chemical shift 6.6 ppm) as a function of diffusion weighting. Top row shows the results for the CPCl/NaSal systems. In the above subfigures, each symbol corresponds to different diffusion time $T_{diff}$; squares (15 ms), diamonds (50 ms), triangles (200 ms) and circles (500 ms). The shaded area, continuous line, dotted line and dashed line correspond to the best fit of Eq. (5), Eq. (6), Eq. (4) and Eq. (3), respectively, and discussed in section III of the paper.

Fig. 6(a) shows a summary of the parameter $\alpha$ as a function of salt to surfactant concentrations for the two WLMs. The approach of $\alpha$ towards linearity mentioned above can be clearly observed in the two systems. It is evident both from Figs. 5 and 6 that each microstructural regime possesses its own diffusion behavior with different $\alpha$ values; linear ($\alpha \approx 0.5$), moderately branched ($0.5 < \alpha < 1$), and the network of branched micellar structures ($\alpha \approx 1$). We note that this analysis was performed for the peak at the chemical shift 6.6 ppm for both CPCl/NaSal and CTAB/NaSal solutions. Similar results are obtained by analyzing other peaks that are associated with the surfactant molecules. These findings illustrate that the diffusion weighted NMR spectroscopy measurements are capable of distinguishing micellar microstructures from each other. Similar diffusive patterns have been reported for reverse micelles based on lecithin in isooctane/cyclohexane via PGSE NMR experiments [26]. Another important aspect of Figs. 5 and 6(a) is that the self-diffusion mechanisms in CTAB/NaSal solutions are similar to that of the CPCl/NaSal solutions over the entire range of salt to surfactant concentration, confirming our hypothesis that the CTAB/NaSal solution should experience a linear to branched micellar transition beyond the first viscosity peak akin to the CPCl/NaSal solution.

Additionally, Fig. 6(b) shows the MSD as a function of the diffusion time for the three micellar regimes noted above. Clearly, the self-diffusion behavior is consistent over the whole range of diffusion times. For moderately branched reverse micelles, Angelico et al. [26] reported a transition from $\alpha \approx 0.5$ to $\alpha \approx 1$ beyond a critical observation time in some of their branched reverse micellar solutions. A similar transition is not reported in our experiments with moderately branched micellar solutions. Perhaps, the distance between branches in our micellar solutions is small such that even for the lowest observation times, the self-diffusing surfactants feel the effects of branched points.

The results of Fig. 5 also indicate that the attenuation in the NMR signal intensity is different across various micellar microstructures. In linear WLMs, the NMR signal decays gradually until it levels off to an asymptotic value. As micellar branches form, the asymptotic behavior weakens until at high branching densities (*i.e.*, high salt-to-surfactant ratios), the NMR signal decays monoexponentially. These distinct trends imply that the surfactant diffusion mechanism is different across various micellar microstructures.

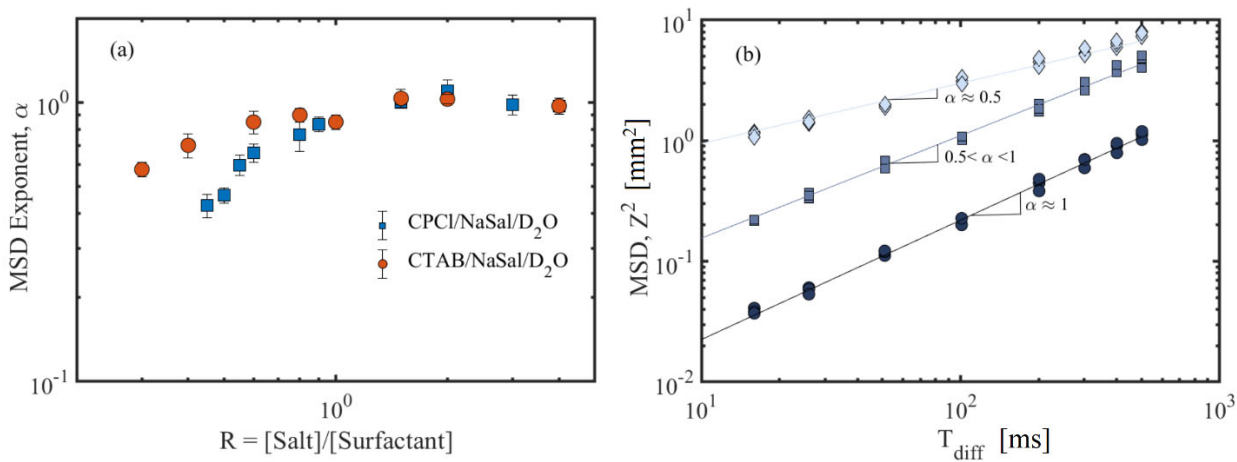

**FIGURE 6.** (a) MSD exponent ($\alpha$) as a function of salt-to-surfactant concentration ratio, R, for the two WLMs. These $\alpha$ values are obtained by superimposing the experimentally measured normalized NMR signals shown in Fig. 5. (b) MSD as a function of diffusion time shown for three self-diffusion behaviors reported in this study. Circles, squares and diamonds correspond to multiple sets of experimental data at high salt-to-surfactant ratios for moderately branched and linear wormlike micelles, respectively. Note that the MSD plots for each micellar regime have been shifted vertically for visual guide. Such vertical shifts do not affect the MSD scaling with respect to the diffusion time.

The next question to address is the mechanism(s) of surfactant diffusion in various micellar microstructures. In order to determine the mechanism of the surfactant self-diffusion in different microstructural regimes, we fitted the experimentally determined NMR signal intensities to appropriate diffusion models noted above in section III, and obtained the ADC over the entire range of salt-to-surfactant concentrations. Then, by monitoring the variation of the self-diffusion coefficients with breakage time ratio and entanglement density, we identified the best fit theory of the self-diffusion in each microstructural regime.

Starting with the linear WLMs (*i.e.*, Fig. 5 (a,d)), it is evident that the NMR signal intensities do not exhibit a mono-exponential decay in this regime. Therefore, Eq. (3) is not appropriate and cannot be fitted to the experimental data. In fact, in agreement with the $\alpha$ observations above and for linear WLMs, the MSD is proportional to the root square of the diffusion time, which is similar to the predictions of the model by Angelico et al. [32]. Therefore, this model is appropriate in this regime, and we have fitted this relation (Eq. 4) to the experimental data to obtain the ADC of the surfactant molecules. The dotted lines in Fig. 5 (a,d) show the results of this fitting process for linear micellar solutions in both systems. It appears that the fit to the experimental data accurately describes the NMR signal at small diffusion weighting, while deviating slightly from the experimental data at higher gradient values. We note that in fitting Eq. (4) to the experimental data, there are two unknown parameters, the apparent curvilinear diffusion coefficient $D_c$ and a characteristic step length, $\lambda$. It has been postulated that $\lambda$ is linked to the persistence length of the micellar chains $l_p$ [32]. However, when the model was fitted with the previously reported $l_p$ values for both micellar systems [13, 37], the resulting ADC were extremely large (~10 cm$^2$/s). Ambrosone et al. [46] noted a similar problem for relating $\lambda$ to a physical property of the micellar system. Therefore, the relation between $\lambda$ and microstructural properties of the micelles still remains unclear and, therefore, poses a new challenge for those who are involved in theoretical modeling of diffusion in self-assembled systems. Consequently, obtaining the apparent curvilinear self-diffusion coefficient of surfactants in micelles is not possible with this method and instead values of $\lambda \sqrt{D_c}$ are reported in Table 1.

An alternative way to quantify the apparent self-diffusion of surfactants in the linear wormlike micelles is to use the theory proposed by Tanner and Stejskal (*i.e.*, Eq. (5) in section III) for restricted diffusion of small molecules in a rectangular channel. The estimated micellar contour length in the linear wormlike micellar solutions (see data in Table 1) are much larger than the size of the individual surfactants. Hence, as a first potential configuration and approximation, we use this Tanner and Stejskal model to quantify surfactant curvilinear self-diffusion inside linear micellar chains. Shaded curves and the confidence intervals around them in Fig. 5(a,d) show the predictions of the best fitted Tanner-Stejskal model to the experimental data of the linear WLMs.

The confidence intervals are necessary due to the superimposition of multiple datasets. For linear wormlike micellar solutions (small R values), this model fits closely to experimental data at low diffusion weighting, but underpredicts the experiments at higher diffusion weighting. The linear WLMs are highly flexible and curved because their contour-length is much larger than their persistence length. Through the Tanner and Stejskal model, we approximated the micellar tubes as rectangular channels. This simplification may have led to deviations with experimental data at high diffusion weightings. To assess the effects of geometry, we fitted the cylindrical model of Callaghan and co-workers (Eq. 6) to the experimental results of linear WLMs. The continuous curves in Fig. 5(a,d) show the best fit of Eq. 6 to the experimental data. The fitted model shows an excellent match with the experimental data, with negligible deviations at high diffusion weightings.

At still higher salt-to-surfactant concentration ratios (and as micellar branches start to form), $\alpha$ increases beyond 0.5 ($\alpha \approx 0.8 \pm 0.1$). Therefore, the model proposed by Angelico et al. is not appropriate, and we have not included such predictions in Fig. 5(b,e) or Table (1). Instead, we have fitted Eq. (5) and Eq. (7) to the experimental data, and it appears that these models provide a close fit to the experimental data over a broad range of diffusion weightings. In particular, the 2D model of Callaghan and co-workers provides a better fit compared to that of the rectangular Tanner and Stejskal model. Perhaps this is not surprising. Our interpretation is that as micellar branches start to form on the backbone of a wormlike micelle chain, individual surfactants may self-diffuse through short branches, which in the framework of the wormlike chain can be thought of a 2D self-diffusion. For larger salt-to-surfactant concentration ratios (*i.e.*, R ≥ 1), the NMR signal follows a mono-exponential behavior. In this regime, we have fitted the mono-exponential function (Eq. (3)) to the experimental data. In summary, the general pattern observed in NMR signal attenuation across various micellar structures is well captured by the theoretical model of Callaghan and co-workers [35].

**TABLE 1.** List of viscoelastic surfactant fluids with their properties. As noted in the text, for samples with the MSD exponent different from 0.5, $\lambda \sqrt{D_c}$ are not listed. Additionally, for systems that do not follow a single-mode Maxwell model, $\xi$ values are not reported. N/A denotes not available.

| Fluid | Salt Concentration [mM] | $L_c$ [μm] | $N$ | $\lambda \sqrt{D_c}$ [mm$^2$/s$^{1/2}$] | $\xi$ |
|---|---|---|---|---|---|
| | 30 | 4 | 56 | $5.7 \times 10^{-4}$ | $2 \times 10^{-5}$ |
| | 40 | 10.2 | 225 | $1.74 \times 10^{-3}$ | $6 \times 10^{-6}$ |
| | 60 | 9.8 | 152 | -- | $2 \times 10^{-5}$ |
| CTAB/NaSal | 80 | 2.3 | 39 | -- | $10^{-4}$ |
| | 100 | 0.93 [37] | 22 | -- | $10^{-3}$ [37] |
| | 150 | 0.9 | 18 | -- | $10^{-4}$ [37] |

| CPCl/NaSal | 200 | 2.1 | 44 | -- | $7\times10^{-5}$ |
| | 400 | 1.3 | 25 | -- | $2\times10^{-4}$ |
| | 45 | -- | -- | $2.15\times10^{-3}$ | -- |
| | 50 | -- | -- | $2.28\times10^{-3}$ | -- |
| | 55 | 1.68 | 23.4 | $1.12\times10^{-3}$ | $10^{-4}$ |
| | 60 | 3.8 | 58.7 | -- | $6\times10^{-5}$ |
| CPCl/NaSal | 80 | 1.4 [13] | 23.7 [13] | -- | N/A |
| | 90 | 0.6 [13] | 8.7 [13] | -- | N/A |
| | 150 | 1.75 | 33.6 | -- | $3\times10^{-4}$ |
| | 200 | 2.6 | 49.5 | -- | $10^{-4}$ |
| | 300 | 0.75 [13] | 11.2 [13] | -- | $10^{-4}$ |
| | 400 | 0.5 [13] | 7 [13] | -- | N/A |

Following the fitting procedure, we obtained the ADC in each sample over the entire range of salt-to-surfactant concentrations in both WLMs for the best fitted curves. Fig. 7 shows the resulting ADC of the surfactant molecules as a function of $T_{diff}$ in the two WLMs. Across both systems, the ADC is maximum at short diffusion times. As the diffusion time increases, the ADC decreases until it plateaus at $T_{diff} \geq 100$ *ms*. To rationalize this result, we note that the self-diffusion of individual surfactant molecules is restricted in the micellar systems. Therefore, at short diffusion time scales, the surfactant molecules do not diffuse to a large enough extent to be fully affected by the restrictions. As the diffusion time is increased, the surfactant molecules move longer distances and become subject to more restrictions. More restrictions lead to a lower surfactant self-diffusion coefficient. We now focus primarily on these asymptote values at large T_diff, as they more consistently describe the impact of the wormlike micellar structure on the diffusion mechanism.

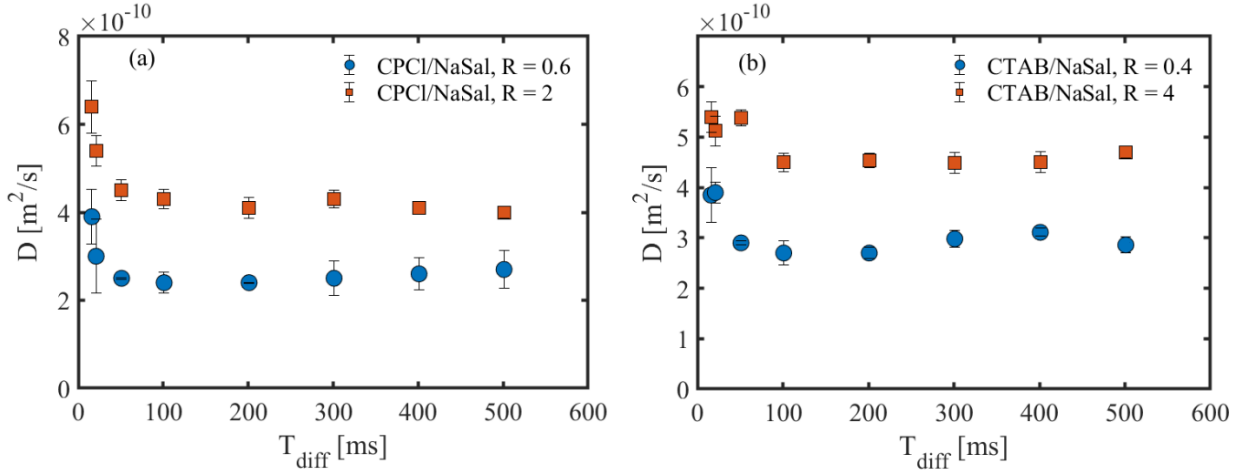

**FIGURE 7.** Surfactant self-diffusion coefficient as a function of diffusion time for different wormlike micellar systems.

Fig. 8 shows the asymptotic ADC of the surfactant molecules as a function of R for the two WLMs used in this study. This figure is a key result of this study, which enables a direct assessment of the primary mechanism of the surfactant self-diffusion in these micellar systems. To discuss the results of Fig. 8, we first start with linear WLMs. In this micellar regime, we first assess the relative importance of the reptation and breakage/reformation on micellar curvilinear diffusion. In CPCl/NaSal system, the first two solutions with R = 0.45 and R = 0.5 are not well described by the single-mode Maxwell model, and therefore, micellar curvilinear diffusion is controlled by a combination of reptation and/or breakage/reformation. As the salt concentration increases, the linear viscoelastic data is well-described by the single-mode Maxwell model, and the micellar breakage time ratio $\xi$ decreases (see Table (1)). A similar trend is observed for the micellar solution of CTAB/NaSal in the linear regime. Therefore, for majority of the linear wormlike micelles, the micellar curvilinear diffusion is controlled by micellar breakage/reformation.

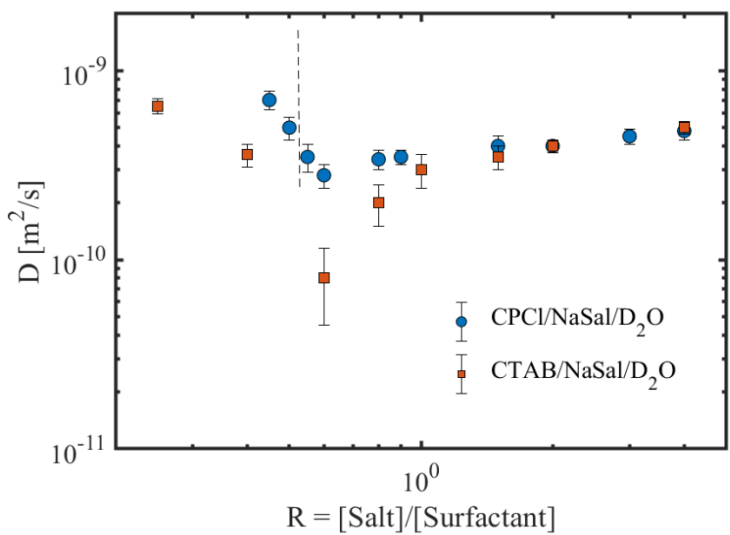

**FIGURE 8.** Apparent self-diffusion coefficient of the surfactant molecules as a function of salt to surfactant concentration ratios for the two wormlike micellar solutions used in this study. The dashed line shows the critical salt to surfactant ratio beyond which the linear viscoelastic response of the CPCl/NaSal solution is best described by a single-model Maxwell model. All CTAB/NaSal solutions are well described by a single-mode Maxwell model.

Now, we turn our attention to the relative significance of micellar curvilinear diffusion to the self-diffusion of individual surfactant molecules. As noted in section (IIA), if the micellar curvilinear diffusion in the fast-breaking regime is the dominant diffusion mechanism, the measured ADC should be inversely proportional to the micelles breakage time ratio $\xi$. Fig. 8 combined with Table (1) indicate that, in the linear WLMs of both systems, a direct proportionality exists with the self-diffusion coefficient decreasing as the micellar breakage time ratio decreases. This trend clearly indicates that the observed diffusion dynamics are not controlled by the micellar curvilinear diffusion. Individual surfactant molecules are much smaller than the micelles themselves. Therefore, they can diffuse extremely rapidly inside the micellar solutions compared to the micelles themselves. As a result, the self-diffusion of individual surfactants should be the dominant diffusion mechanism in micellar solutions. Taken together, not only the effects of micellar curvilinear diffusion on self-diffusion measurements are negligible in the linear WLMs, but also in the branched micellar solutions. Therefore, in the remaining of the discussion, we will only focus on assessing the dominant self-diffusion mechanism of the individual surfactant molecules.

Let us now consider the surfactant self-diffusion in linear micellar solutions. By using a similar argument presented in section II, we note that the surfactant diffusion mechanism due to exchange between micelles is insignificant. For this mechanism to play a dominant role, the ADC should be directly proportional to the micellar entanglement density $N$. Table (1) together with Fig. 8 illustrate that, for linear WLMs, ADC is inversely proportional to $N$, indicating that the diffusion due to surfactant exchange between micelles in linear micellar solutions is negligible. In addition, the diffusion due to exchange between micelles and the continuous phase is not important. The latter diffusion mechanism should produce a MSD that is linearly proportional to the diffusion time. Thus, through elimination of mechanisms of diffusion, we can conclude that the self-diffusion of surfactant molecules in the linear wormlike micelles is strongly controlled by the curvilinear type diffusion of individual surfactant molecules along the contour length of the micellar tubes. The above measurements on linear wormlike micelles also indicate that the surfactant curvilinear diffusion coefficient is inversely proportional to the equilibrium micelles length ($L_c$). Therefore, diffusion-weighted NMR spectroscopy could be used as an alternative technique (besides rheology, cryo-TEM and small angle neutron scattering) to assess the variation of the micellar length in the linear micellar solutions.

At higher surfactant concentrations and past the first viscosity maxima, the ADC of the surfactants increases for both systems as micellar branches form. Using a similar argument as for linear wormlike micellar solutions (*i.e.*, ADC is roughly proportional to $N$), the dominant mechanism of diffusion would be curvilinear with individual surfactant molecules diffusing along the contour length of the micellar chain. On the scale of micelles, surfactant molecules are still diffusing along the micellar longest dimension. But, on a scale of several micelles, they move on random-like patterns due to presence of micellar branches to which they are bound. Therefore, as salt concentration increases, diffusion increasingly resembles the Brownian random-walk behavior with α approaching unity.

As the salt concentration increases beyond R > 0.9 for CPCl/NaSal and R > 1 for CTAB/NaSal ADC still continuously increases. According to rheological measurements, the effective micellar entanglement density increases in this salt concentration range, and therefore, the surfactant self-diffusion due exchange between micelles and/or the bulk are all relevant. However, beyond the second viscosity maxima, as the salt concentration increases, the surfactant self-diffusion coefficient increases. According to the rheological measurements listed in Table (1), micellar entanglement density decreases in this regime. Thus, similar to the above discussion on linear WLMs, the most dominant mechanism of surfactant self-diffusion is the curvilinear diffusion along the contour length of the branched micelles. In this regime, surfactant molecules diffuse curvilinearly on the scale of the individual micelles, while on the scale of a dense network of branched micelles, this diffusion follows a random-like pattern giving rise to a MSD that is linearly changing with diffusion time. Finally, the surfactant ADC in wormlike micelles are much smaller than the diffusion of residual protons in the bulk, which confirms the highly restricted nature of surfactant self-diffusion in these micellar systems.

## VII.    CONCLUSIONS

Harking back to the motivation of this study, we investigated the self-diffusion of surfactant molecules in two model wormlike micellar solutions based on CPCl/NaSal and CTAB/NaSal over a wider range of salt to surfactant concentration ratios. The results of this study can be summarized as follows:

For salt-to-surfactant concentrations below the first viscosity peak, the MSD varies with diffusion time as $Z^2(T_{diff}) \propto T_{diff}^\alpha$ with $\alpha \approx 0.5$. At concentrations beyond the first viscosity peak, the mechanism of surfactant diffusion changes towards a random-walk with $0.5 < \alpha < 1$ with $\alpha$ becoming unity for salt concentrations around the second viscosity peak. The variation of MSD with diffusion time can be used as evidence for formation of micellar branches that generate random patterns. Additionally, we showed that the NMR signal attenuation behavior varies across different micellar topologies. In the linear micelles, the signal gradually decays and shows an asymptotic behavior at high diffusion weightings. As micellar branches start forming on the main backbone, the asymptotic behavior weakens, and eventually at high branching densities (*i.e.*, large salt-to-surfactant ratios), the NMR signal follows a mono-exponential decay. This trend is similar to theoretical predictions of Callaghan and co-workers when diffusion changes from 1D to 2D and 3D assuming a cylindrical pore geometry. Therefore, based on these results we conclude that DW-STEAM NMR is sensitive to the type of micellar microstructure (whether linear or branched).

In addition, the measured ADCs for surfactant molecules in different microstructural regimes are much smaller than the diffusion of proton in the bulk $D_2O$. This result indicates a highly restricted nature of surfactant self-diffusion in micellar systems. We also identified the mechanism(s) of surfactant self-diffusion in these micellar systems. For the linear and moderately branched WLMs (*i.e.*, R < 1), the diffusion mechanisms due to surfactant exchange with the bulk or at entanglement points are negligible, and the most dominant mechanism of surfactant diffusion is curvilinear. As micellar branches form a dense network (*i.e.*, 1< R< 2), the surfactant self-diffusion arises from a combination of surfactant exchange with the bulk and other micelles. Finally, past the second viscosity maxima, the surfactant self-diffusion occurs predominantly through curvilinear diffusion along the micellar contour-length in both micellar solutions.

## SUPPORTING INFORMATION

See the supplementary materials for more details on the small amplitude oscillatory shear data of the WLMs and a table that includes the numerical values of the relaxation time, micellar breakage time and the zero-shear rate viscosity of the wormlike micelles.

## ACKNOWLEDGEMENTS

Part of this work was performed at the US National High Magnetic Field Laboratory (NHMFL), which is supported by the State of Florida and the National Science Foundation Cooperative Agreement No. DMR-1644779. In addition, HM gratefully acknowledges support by National Science Foundation through award CAREER CBET 1942150. SCG and SWH are supported by the US National Institutes of Health through award R01-NS072497. We also acknowledge the reviewers for their insightful comments that improved the quality of the paper.

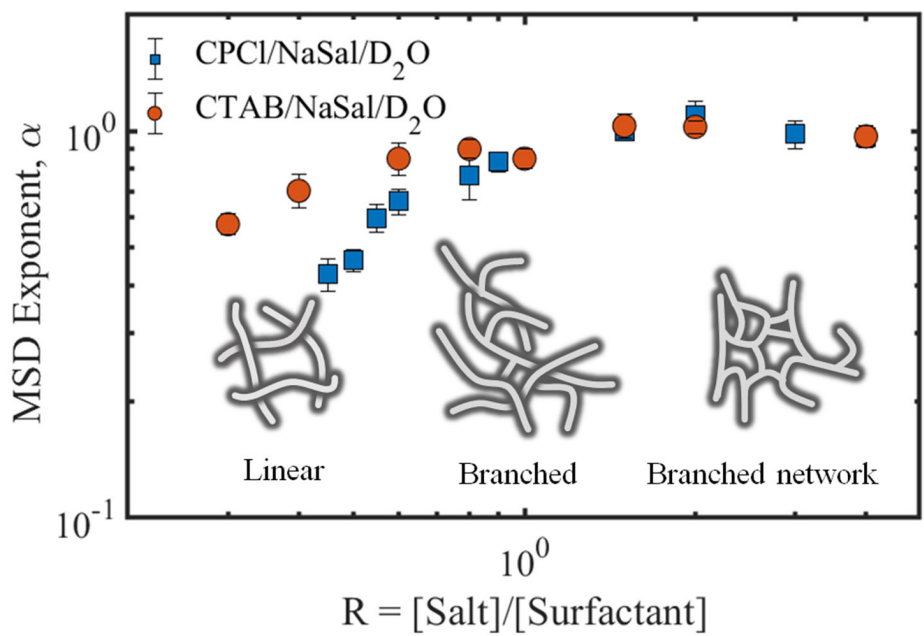

Table of Contents Graphics.